\documentclass[cits]{PoS}

\usepackage{subfigure}

\usepackage{url}

\title{A turn-key Concept for active cancellation of Global Positioning System L3 Signal}

\ShortTitle{A turn-key Concept for active cancellation of Global Positioning System L3 Signal}

\author{\speaker{Lou Nigra}\\
        University of Wisconsin - Madison Department of Astronomy / NAIC\\
        E-mail: \email{nigra@astro.wisc.edu}\\
        Web: \url{http://www.astro.wisc.edu/~nigra/rfi}}
        
\author{B. Murray Lewis\\
        National Astronomy and Ionosphere Center, Arecibo, Puerto Rico\\
        E-mail: \email{blewis@naic.edu}}

\author{Clyde Edgar\\
        The Aerospace Corporation\\
        E-mail: \email{clyde.e.edgar@aero.org}}

\author{Phil Perillat, Luis Quintero,\\
\llap{}National Astronomy and Ionosphere Center, Arecibo, Puerto Rico\\
E-mail: \email{phil@naic.edu}, \email{lquintero@naic.edu}}
\author{Sne\v{z}ana Stanimirovi{\'{c}} and J. S. Gallagher, III\\
\llap{}University of Wisconsin, Madison Department of Astronomy\\
E-Mail: \email{sstanimi@astro.wisc.edu}, \email{jsg@astro.wisc.edu}}

\abstract{We present a concept, developed at the National Astronomy and Ionosphere Center (NAIC) at Arecibo, Puerto Rico, for active suppression of Global Positioning System (GPS) signals in the 305 m dish radio receiver path prior to backend processing. The subsystem does not require an auxiliary antenna and is intended for easy integration with radio telescope systems with a goal of being a turnkey addition to virtually any facility. Working with actual sampled signal data, we have focused on the detection and cancellation of the GPS L3 signal at 1381.05 MHz which, during periodic test modes and particularly during system-wide tests, interfere with observations of objects in a range of redshifts that includes the Coma supercluster, for example. This signal can dynamically change modulation modes and our scheme is capable of detecting these changes and applying cancellation or sending a blanking signal, as appropriate.
The subsystem can also be adapted to GPS L1 (1575.42 MHz), L2C (1227.6 MHz), and others.
A follow-up is underway to develop a prototype to deploy and evaluate at NAIC.}

\FullConference{RFI mitigation workshop\\
		 29-31 March 2010\\
		 Groningen, the Netherlands}

\begin{document}

\section{Introduction}

Navigation satellite systems, such as Global Positioning System (GPS) and the Russian \emph{Globalìnaya Navigatsionnaya
Sputnikovaya Sistema (GLONASS)}, are inherently troublesome to radio astronomy. Ground-based Radio Frequency Interference (RFI) typically enters the system from near the horizon through antenna sidelobes with gain $\sim0$~dBi. Satellite signals, on the other hand, can arrive at any angle through  sidelobes with much higher gain. In this work, we describe a concept for mitigating the troublesome GPS signal at L3, which frequently interferes with the operation of neutral Hydrogen (HI) surveys at the National Astronomy and Ionosphere Center's Arecibo observatory (AO), in Puerto Rico.

GPS satellites currently transmit on three carrier frequencies. L1 (1575.42 MHz) carries the well-known civilian signal and a more sophisticated encrypted signal for military purposes. L2 (1227.6 MHz) also carries an encrypted military signal, but a planned civilian signal is currently being tested. L3 (1381.05~MHz) carries only a non-navigation signal with encrypted data, used as part of a nuclear event detection system. All three of these carriers employ Binary Phase Shift Keying (BPSK) modulation where a pre-defined Pseudo-Random Noise (PRN) sequence modulates data at a lower rate, which carries the system information. This spread-spectrum baseband signal, in turn, modulates the carrier. The L1 civil signal and L3 both use the same 1.023 Mb/s PRN code (unique to each satellite) which repeats every 1~ms. The L1 data rate is 50~bps, but the L3 data rate is notably much higher at 11~kbps. The resulting $\sin^2(x)$~/~$x^2$ power spectra of these signals have width of $\sim1$~MHz, or $220$~km~sec$^{-1}$. Received signal levels are fairly constant, nominally $-127$~dBm (3~dBi antenna) \cite{gps200d}, equivalent to a $1.3\times10^{6}$~Jy source.


Although L1 is on continuously, L3 is sporadic. Several times daily, satellites undergo tests and transmit for seconds at a time. System-wide tests are performed quarterly in windows lasting for days during which satellites transmit for minutes at a time. The occurrence and length of transmissions are unpredictable. The authorities coordinate with the radio astronomy community by giving advance notice of these test windows. While transmitting, the signal is generally unmodulated by data (PRN modulation alone), but will periodically overlay its unique 11~kbps data for a number of 1.5~sec slots. L3 is a particular problem because its spectrum overlaps red-shifted 21~cm HI emission (1420.4 MHz at rest) from the Coma galaxy supercluster at $z=0.03$. Making matters worse, its $220$~km~sec$^{-1}$ wide spectrum is similar to that of spiral galaxy signatures. Currently, the strategy at AO is to flag data when daily tests corrupt observations and the band is avoided during announced system test windows resulting in shutting down surveys for days at a time.

\section{Cancelation Techniques}

\begin{figure}
\centering
\subfigure[Method of \cite{barnbaum98}, \cite{poulsen05} using reference antenna.]{
\includegraphics[width=0.45\textwidth]{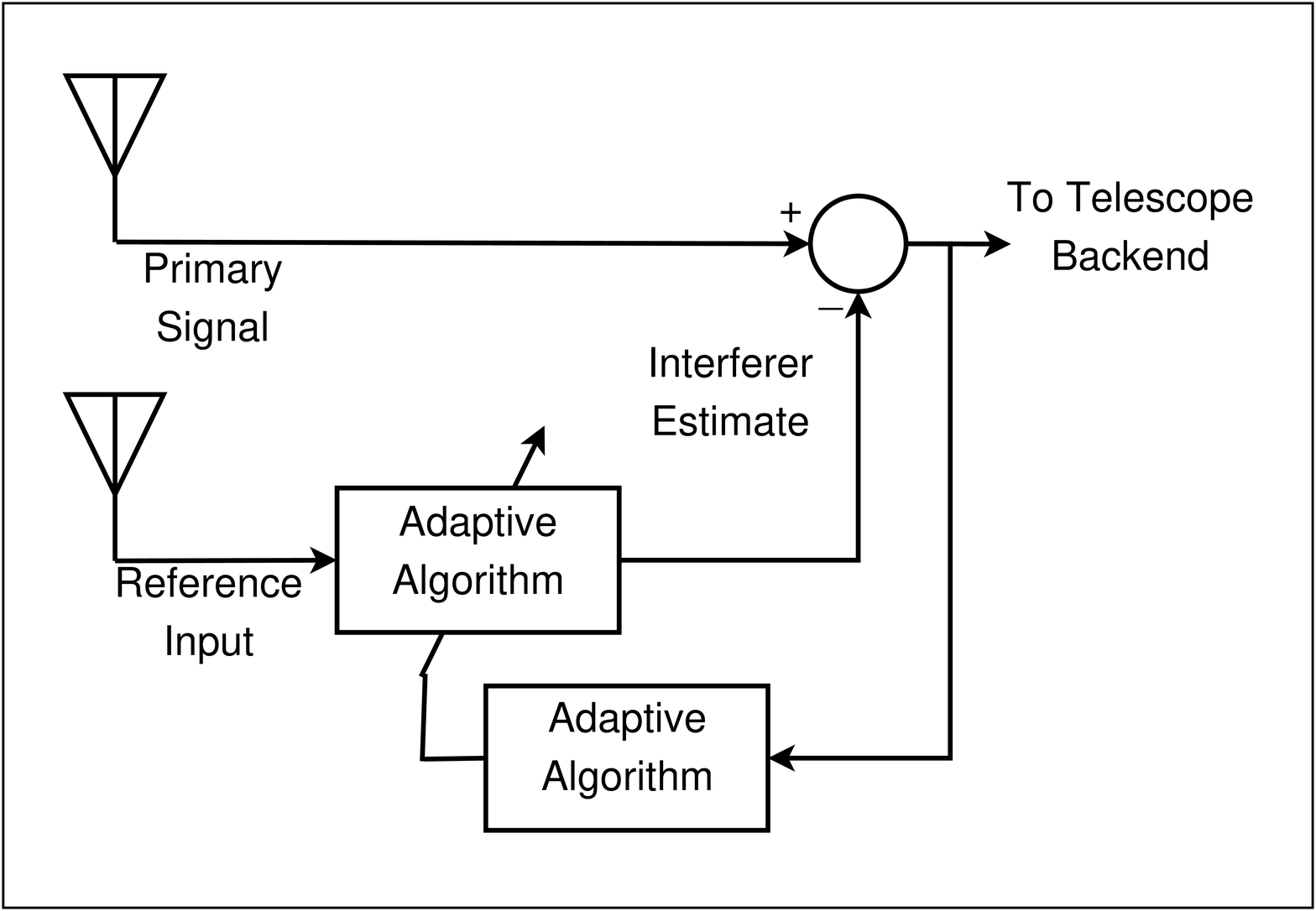}
\label{adaptivefilter}
}
\subfigure[Method of \cite{ellingson01} using no reference antenna.]{
\includegraphics[width=0.35\textwidth]{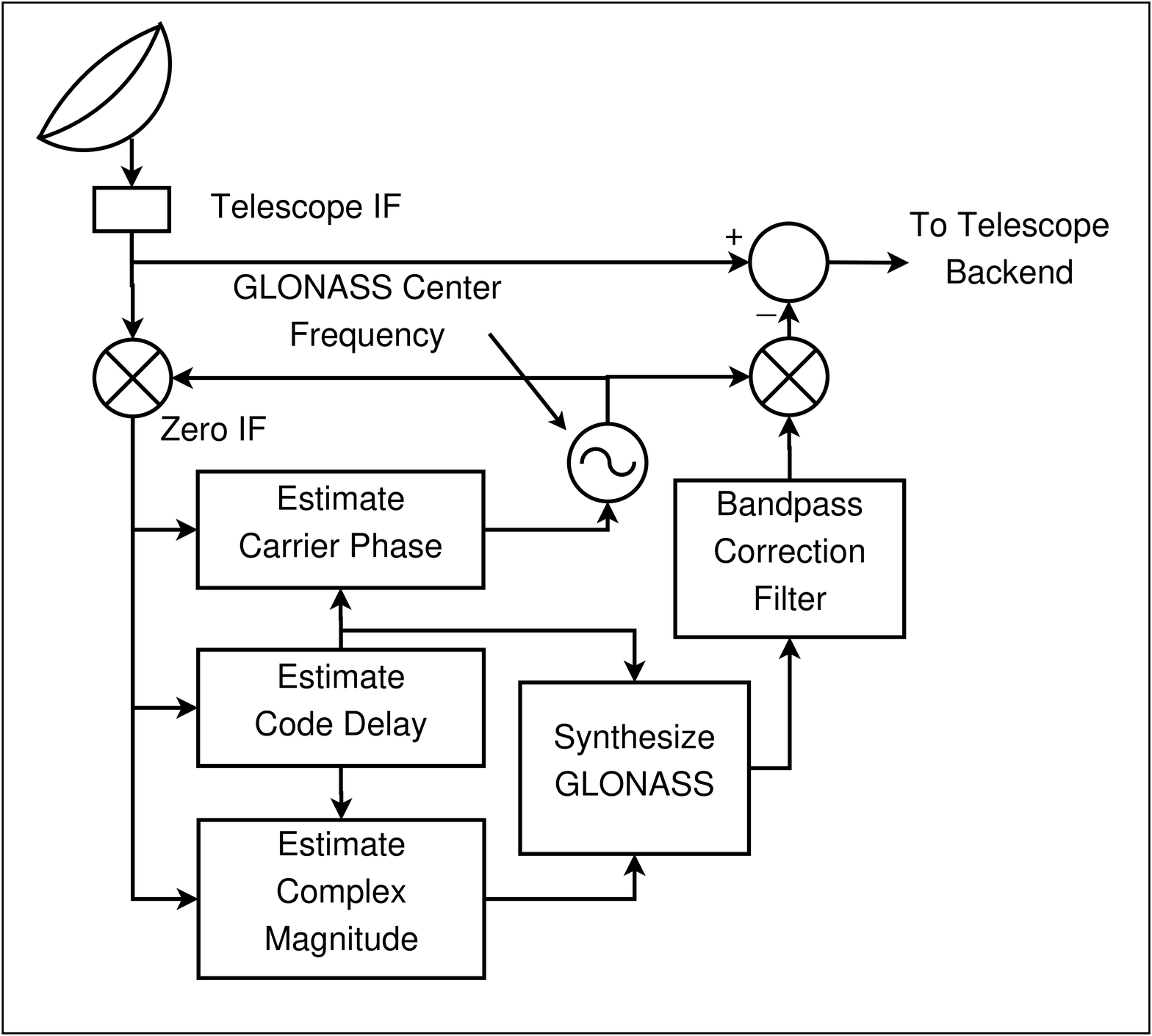}
\label{ellingson}
}
\label{cancelationmethods}
\caption[Optional caption for list of figures]{Prior approaches to adaptive cancellation of RFI in radio telescopes. See text for explanation.}
\end{figure}

The adaptive filter has a long history in electronic systems and is well-suited for canceling a time-varying, unwanted signal. Figure \ref{adaptivefilter} shows a time domain adaptive filter in a RFI canceling configuration. A high quality replica of the transmitted interferer is required as input and its output is subtracted from the system channel. An algorithm adapts the filter's characteristics to optimally match the channel response by minimizing the interferer after subtraction. In \cite{barnbaum98}, suppression of commercial FM signals with such an adaptive filter was successfully demonstrated. Relevant to the problem at hand, in \cite{poulsen05} a similar technique was applied to navigation satellite RFI, successfully suppressing a GLONASS signal. In both of these implementations, a separate antenna is required to obtain a reference signal with much higher signal-to-noise ratio (SNR) than the interfererence. In the latter case (\cite{poulsen05}), a tracking dish is required to track the satellite. Given that 4 to 11 GPS satellites are in the sky at a given time, it is likely that several satellites produce strong interference at any given time from different directions. In fact, we have detected three GPS satellites in a single 30-second sample signal from the AO telescope. Using the method of \cite{poulsen05} would require as many tracking dishes as the number of satellites, making it impractical and is clearly inconsistent with our "turn-key" concept.

The need for reference antennas is eliminated for a class of spread-spectrum interferers with high \emph{processing gain} (the ratio of PRN to data bandwidths) by a clever approach described in \cite{ellingson01}, and shown in Figure \ref{ellingson}.  It exploits the fact that most of the bandwidth of spread-spectrum systems such as GPS and GLONASS is redundant, produced by the pre-defined PRN waveform. In this scheme, the reference signal is obtained from the telescope signal channel itself, the PRN modulation is removed, and the signal is filtered to a lower bandwidth, thus increasing its SNR. This is then noiselessly re-expanded with the PRN, producing the necessary high SNR reference signal. In this case, a frequency domain adaptive filter implementation is used rather than time domain, but the principles are similar. A limitation of the approach is that with narrower reference bandwidth, group delay in the reference signal filter increasingly de-correlates the replica signal with the interferer, limiting the SNR improvement to about 1/10 of the processing gain. In the case of standard GPS (L1), SNR improvement by a factor of 2000 can be obtained, but only by a factor of 9 for L3 with its high data rate. Still, this approach eliminates the multiple antenna requirement and with improvements, serves as the basis for our turn-key subsystem.

\section{A Turn-Key GPS RFI Suppression System Concept}

Figure \ref{system} shows our RFI suppression subsystem concept, which can simply be inserted into the Intermediate Frequency (IF) path of a radio telescope system just prior to the backend and can be implemented rather simply with Field Programmable Gate Arrays (FPGA). A fully functional MATLAB$^{\textregistered}$ object-oriented simulation  of the subsystem has been developed. Currently with only a single cancellation channel, it is capable of processing actual data captured from the telescope and multi-channel capability is being added.
The analog IF signals require digital converters that must operate at a minimum rate on the order of the IF bandwidth, currently 300 MHz at AO. Only the frequency conversions and the delay line memory operate at this rate, well within modern device capabilities. Virtually all processing occurs at much lower rates. The processing capacity requirements of the subsystem haven't been determined yet, but this is limited only by the number and type of FPGA's employed.
The basic concept is similar to that in \cite{ellingson01} in that the telescope signal is down-converted, the interferer is detected and modeled, then applied to an adaptive filter and subtracted from the telescope signal, but there are important differences:
%
(1) A digital, programmable delay of the telescope IF signal is included which eliminates delay de-correlation limitations. It can be implemented with $\sim80$~MB of RAM for $0-20$~ms, adequate for the demodulation delays of GPS L1, L2C and GLONASS.
(2) A multi-channel GPS receiver\footnotemark \footnotetext{The concept of including a standard receiver is suggested in \cite{ellingson01}.} is incorporated featuring a unique L3 detection mode and is extensible to other signals such as GPS L1, L2C and GLONASS.

Demodulation of the signal by the receiver and delay compensation allows maximum reference signal quality and therefore maximum suppression to be achieved. Aside from set up commands, the subsystem can run autonomously, with the receiver detecting and  tracking interferers and automatically applying the appropriate mitigation strategy. Blanking would be used while the adaptive filter settles to new signal conditions detected by the receiver, such as when the L3 signal switches between modulated and unmodulated modes.

\subsection{L3 Detection/Demodulation}
\begin{figure}
\begin{minipage}[b]{0.45\linewidth}
\centering
\includegraphics[width=1\textwidth]{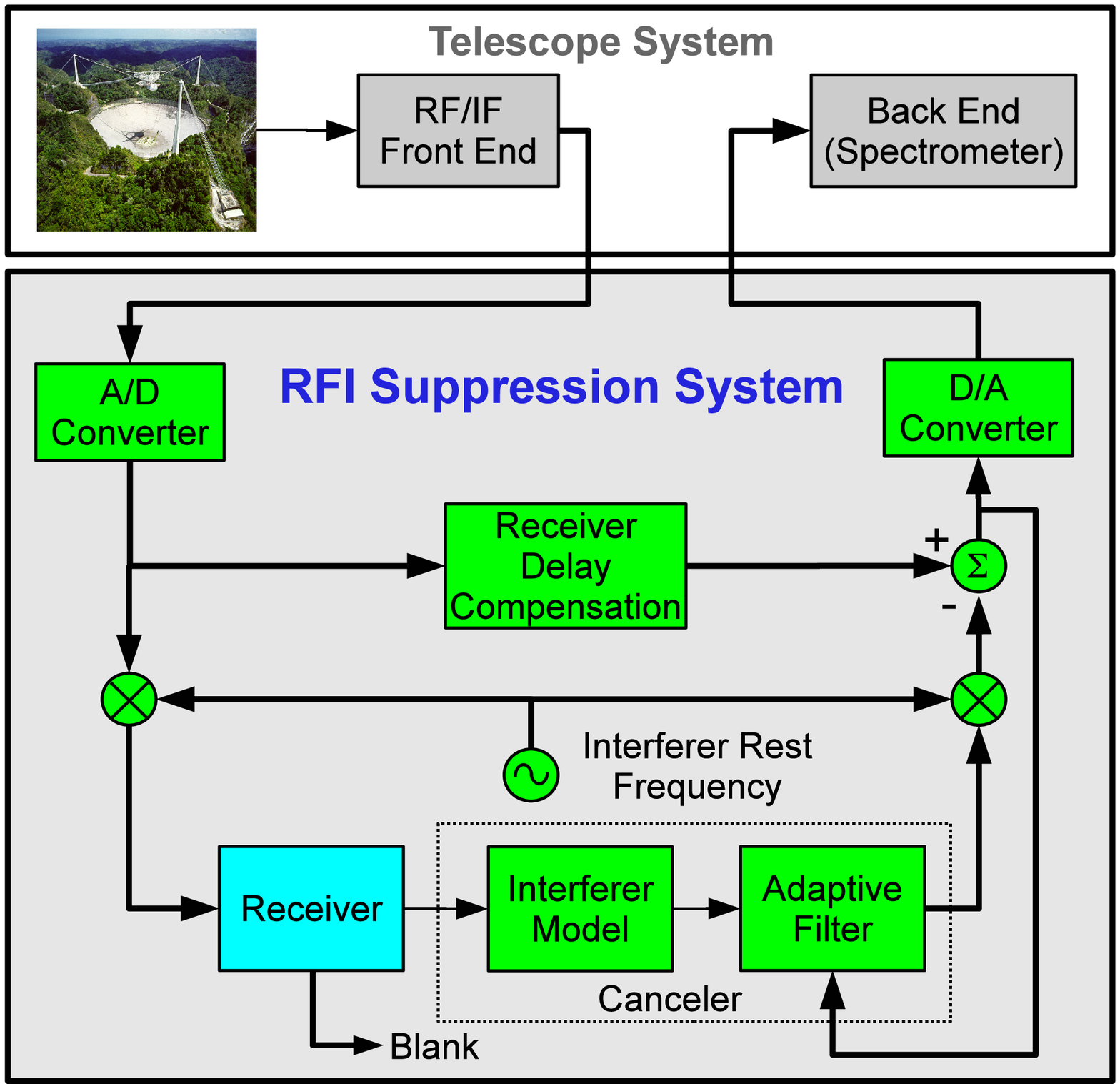}
\caption{A turn-key autonomous subsystem concept for RFI suppression inserted in the radio telescope IF. See text for explanation}
\label{system}
\end{minipage}
\hspace{0.05\linewidth}
\begin{minipage}[b]{0.50\linewidth}
\centering
\includegraphics[width=1\textwidth]{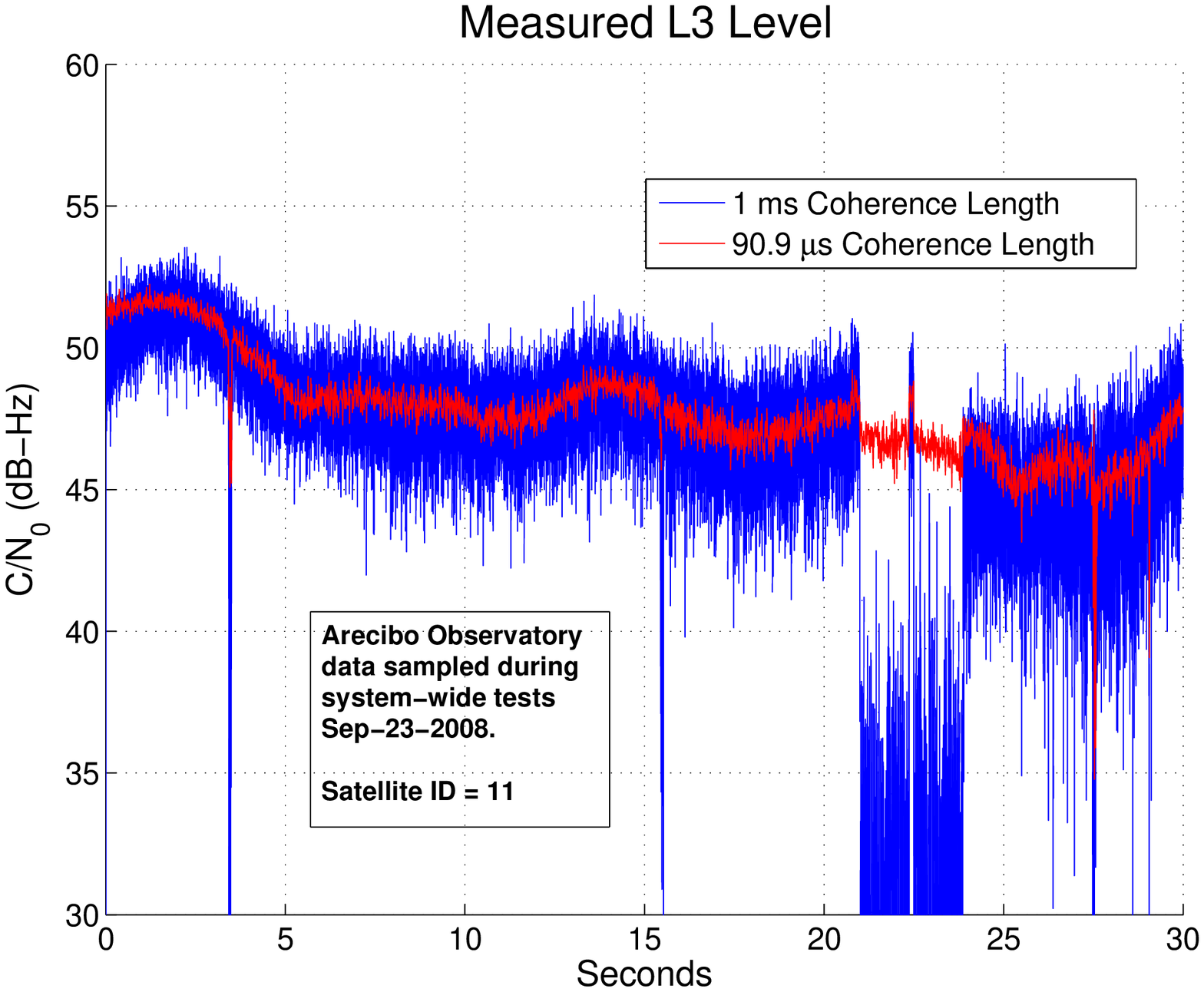}
\caption{Measured level of a L3 signal using two methods over 30 seconds of actual captured AO telescope data. The 1 ms coherence method (blue) drops out when the L3 signal is modulated, but the 90.9 $\mu$s coherence method does not. See text for explanation.}
\label{L3detection}
\end{minipage}
\end{figure}

Our unique method of detecting the L3 modulation mode is based on comparing the amplitude estimates using 1 ms coherent integrations of the correlated signal to an incoherent integration (power sum) of 121 sequential 90.9~$\mu$s coherent integrations. The latter is the power sum of 121 L3 data bits. Each data bit has an SNR that is 11 times lower than the 1 ms coherent integration, but the power sum yields an improvement of $121^{0.5}=11$ times, giving the two measurements equal SNR and thus the same detection sensitivity when no modulation is present. However, when the L3 modulation kicks in, the coherence length drops from 1 ms to 90.9~$\mu$s and the 1 ms measurement drops dramatically, while the other measurement remains unchanged. This detection method has been demonstrated with sampled data taken during one of the L3 system-wide tests. The plot of Figure \ref{L3detection} shows the measured signal levels by the two methods over a 30~s period. Note that at 21~s, the 1~ms integration measurement drops out for 3~s with a brief return midway. This is the satellite going into its 1.5~s slotted modulation mode for two time slots and then returning to unmodulated.

When unmodulated, suppression can be maximized since the entire signal is completely known. When modulated, the high data rate yields an unusable SNR when demodulated on a single bit basis, so rather than cancel, blanking for several seconds during modulation could be a reasonable strategy for some observation modes, such as tracking. At AO however, meridian nodding is the preferred mapping mode and lengthy blanking is not desirable. As such, we are investigating methods of demodulating L3 data in long blocks of bits to achieve a SNR comparable to that of standard GPS so that cancellation is feasible for L3.

\subsection{Canceler}

\begin{figure}
\begin{minipage}[b]{0.5\linewidth}
	\centering
	\includegraphics[width=1.0\textwidth]{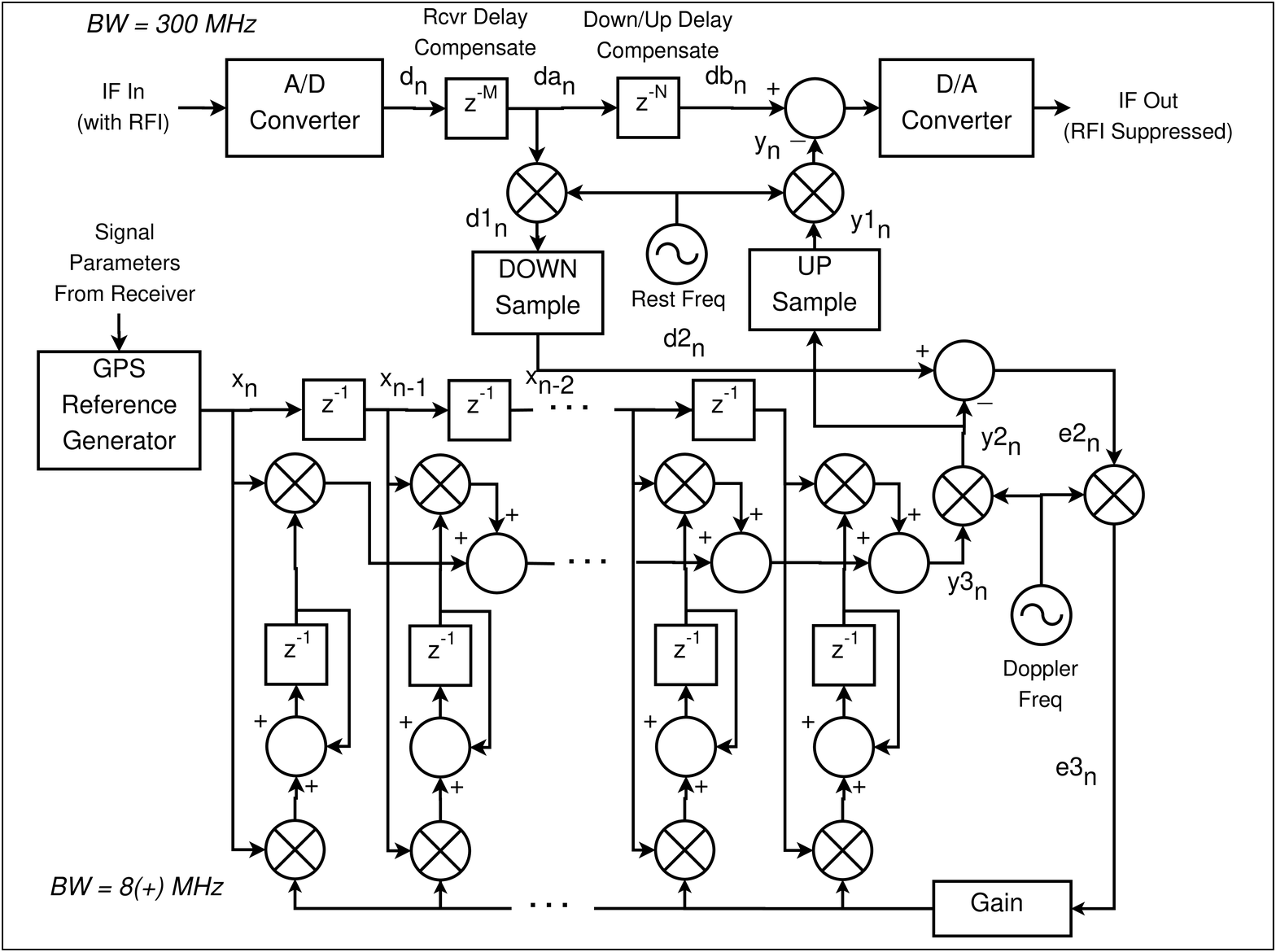}
	\caption{Canceler and its IF interface. See text for explanation.}
	\label{canceler}
\end{minipage}
\hspace{0.03\linewidth}
\begin{minipage}[b]{0.5\linewidth}
	\subfigure{
	\includegraphics[width=0.8\textwidth]{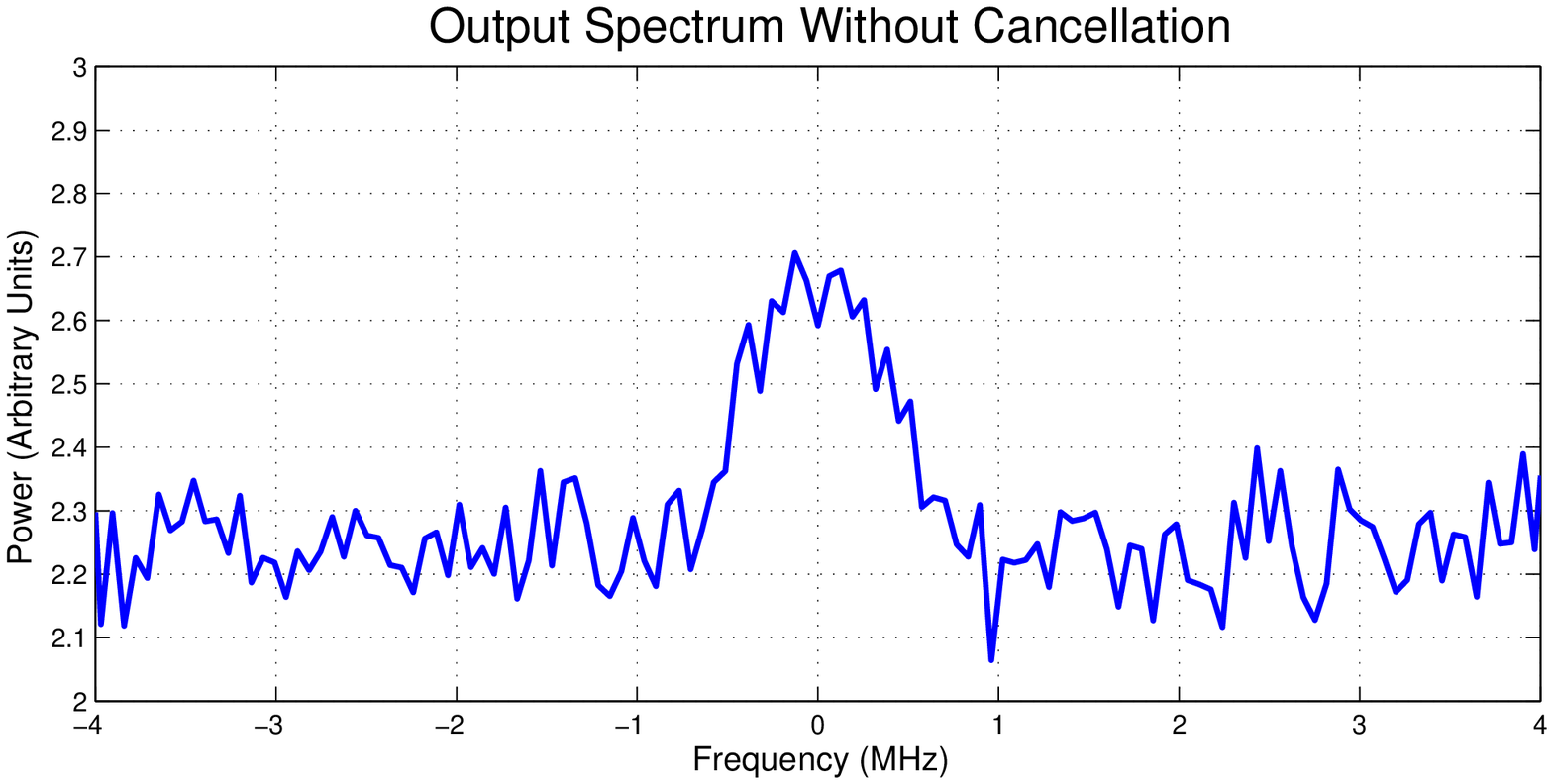}
	\label{syssimnocancel}
	}
	\subfigure{
	\includegraphics[width=0.8\textwidth]{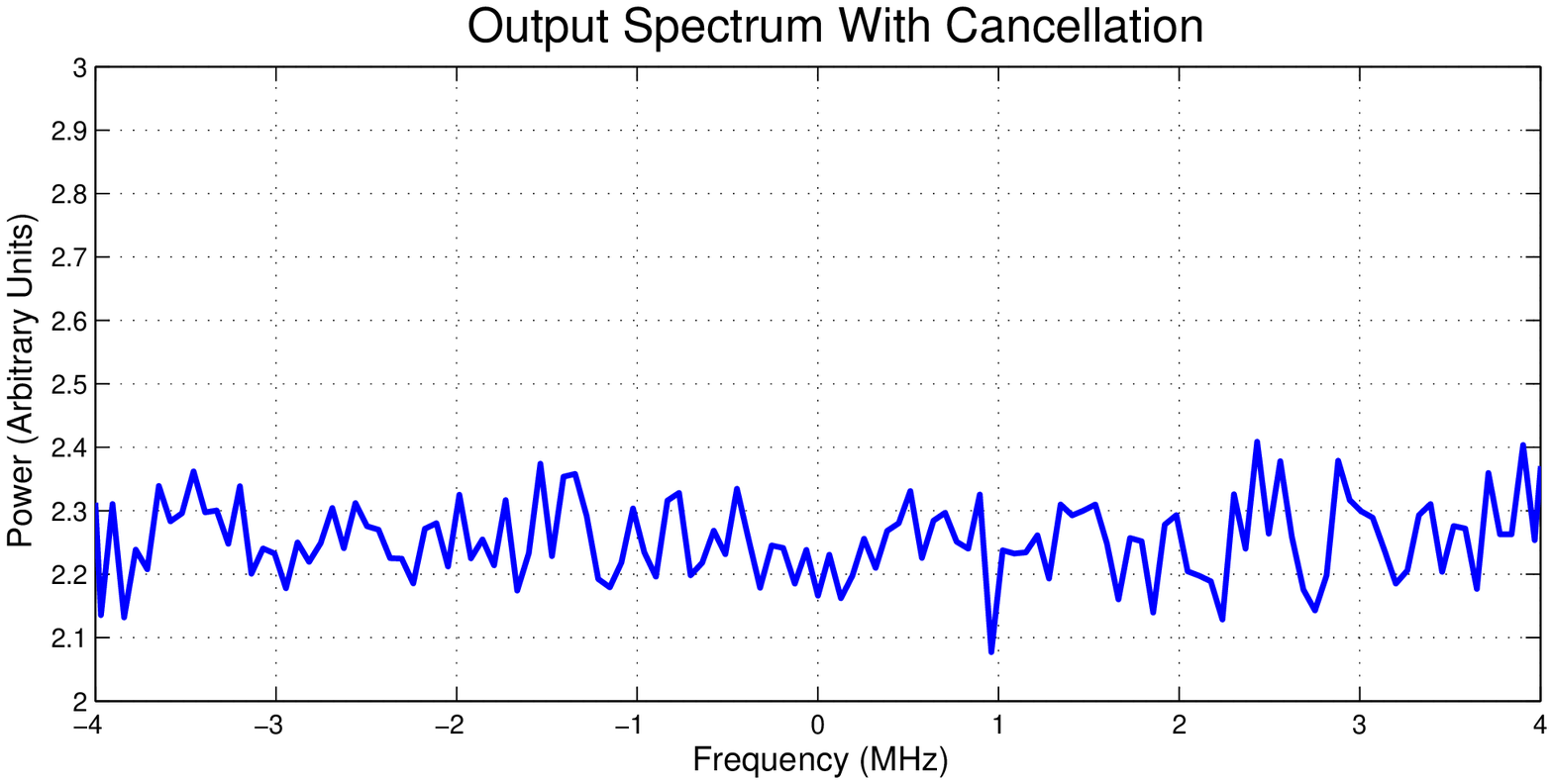}
	\label{syssimcancel}
	}
	\caption[Optional caption for list of figures]{Simulated system response without (top) and with
	(bottom) cancellation. See text for explanation. }
	\label{syssim}
\end{minipage}
\end{figure}

In Figure \ref{system}, the Interferer Model and Adaptive Filter blocks represent the  canceler subsystem illustrated in the lower part of Figure \ref{canceler}. There, the adaptive filter structure is shown along with the two frequency shifts; a fixed IF offset, and a variable doppler offset (maintained by the receiver), required to allow the reference generator and adaptive filter to operate at zero IF. Figure \ref{system} is simplified to show only one canceler but in practice, there will be one for each satellite signal detected. The canceler performance is being evaluated as a separate component as well as in the complete subsystem simulation. Figure \ref{syssim} shows initial simulated spectrometer results with and without cancelation enabled for the single channel subsystem simulator with a simulated telescope signal input. The signal is composed of system noise and a single GPS L3 interferer with no modulation at a level of C/N$_0=52$~dB-Hz, equal to the highest level detected in our sampled telescope data. The spectra represent an observation with 40~ms of integration and resolution of 0.8~km~sec$^{-1}$, smoothed to 13~km~sec$^{-1}$. The canceler leaves no apparent residual RFI at this integration time, but astronomical observations integrate for much longer than this. We are developing simulations that allow us to probe the inner workings and internal signals of the canceler and allow us to predict the performance for actual astronomical observations, with integrations times in tens of seconds. 

\section{Summary and Future Plans}

We are developing a GPS RFI suppression subsystem concept specifically for GPS L3, but extensible to other navigation satellite signals. The subsystem, requiring no external antennas or other hardware, is designed to easily insert into a radio telescope system and operate virtually autonomously requiring a minimum of setup, control and coordination from the system. It is simulated with MATLAB$^{\textregistered}$ object-oriented code facilitating the transition to a FPGA-based prototype. A unique method of detecting transitions between GPS L3's two modulation modes has been demonstrated and will be used to automatically adjust the mitigation technique. Another unique feature is a programmable delay in the path of the system signal which enables maximum cancellation potential, including possible methods for demodulation and cancelation of the difficult, high data rate mode of L3. The simulator is fully functional, able to cancel a single simulated satellite signal and the capability for canceling multiple signals is nearly available. The cancelation method is being evaluated and optimized for long, realistic astronomical observations.

We will develop more sophisticated simulated interferers with realistic signal dynamics and distortion based on the analysis of our receiver tracking response to actual captured AO data. Of course, with multiple canceler channel capability added to the subsystem, we will be able to demonstrate performance with actual captured AO data. The canceler is currently based on the simple LMS algorithm and we will evaluate other, more complex algorithms that should achieve better performance. The simulator will be modified to specifically accommodate GPS L1, L2C and GLONASS signals. A functional FPGA-based prototype is planned for development at AO. Additional related "spin-off" projects are being considered: (1) Develop a GPS situation display for the observer that would graphically identify satellite locations, 
indicate mitigation status, and the potential for interference based on location and nominal antenna response. (2)
Investigate the feasibility of logging a spectral model of the residual RFI based on receiver and canceler responses.
This would be used by the observer to model and remove residual structure from the baseline.


\end{document}